\documentclass[a4paper]{article}

\usepackage{ISCSLP2022}

\usepackage{graphicx}
\usepackage{tabularx}
\usepackage{amsmath,amssymb,amsthm}
\usepackage{booktabs}

\title{Speech Enhancement Based on CycleGAN with Noise-informed Training}

\name{Wen-Yuan Ting$^{1}$, Syu-Siang Wang$^{2}$, Hsin-Li Chang$^{3}$, Borching Su$^{1}$ and Yu Tsao$^{4}$}
\address{$^{1}$Graduate Institute of Communication Engineering, National Taiwan University, Taipei, Taiwan\\
	$^{2}$Department of Electrical Engineering, Yuan Ze University, Taoyuan, Taiwan\\
	$^{3}$Department of Electrical Engineering, National Central University, Taoyuan, Taiwan\\
	$^{4}$Research Center for Information Technology Innovation, Academia Sinica, Taipei, Taiwan}
\email{}
\begin{document}
	\ninept
	\maketitle
	\begin{abstract}
		Cycle-consistent generative adversarial networks (CycleGAN) were successfully applied to speech enhancement (SE) tasks with unpaired noisy--clean training data. The CycleGAN SE system adopted two generators and two discriminators trained with losses from noisy--to--clean and clean--to--noisy conversions. CycleGAN showed promising results for numerous SE tasks. Herein, we investigate a potential limitation of the clean--to--noisy conversion part and propose a novel noise-informed training (NIT) approach to improve the performance of the original CycleGAN SE system. The main idea of the NIT approach is to incorporate target domain information for clean--to--noisy conversion to facilitate a better training procedure. The experimental results confirmed that the proposed NIT approach improved the generalization capability of the original CycleGAN SE system with a notable margin.
	\end{abstract}
	
	\section{Introduction}\label{sec:intro}
	Owing to recent advances in machine-learning techniques, speech-related applications have been widely deployed in our daily lives. However, in real-world scenarios, speech signals are disturbed by environmental noise, resulting in poor voice quality and low intelligibility, which limits the performance of downstream tasks. To address this issue, numerous speech enhancement (SE) techniques have been deployed as preprocessors to convert noisy speech signals to clean ones \cite{chao2021cross,hartmann2013direct}.
	
	Based on the availability of paired noisy-clean speech signals during training, SE approaches are divided into two categories requiring different types of data: (1) paired noisy--clean training data and (2) unpaired noisy--clean training data \cite{mohammadiha2013supervised,kim2018unpaired,neri2021unsupervised,venkataramani2019style}. Approaches in the first category are referred to as ``supervised SE methods." These methods learn a mapping function that characterizes the noisy--clean transformation. At runtime, the mapping function is used to perform denoising. Recently, deep neural networks \cite{zhao2018two,xu2014dynamic}, which have shown the strong capability for modeling nonlinear transformations, have been used to form the mapping function for SE and achieve state-of-the-art enhancement performance. Well-known examples include deep denoising autoencoders \cite{lu2013speech}, convolutional neural networks \cite{pandey2019tcnn}, long-short term memory \cite{sun2017multiple}, and transformers \cite{kim2020t}. Identifying a suitable objective for training the mapping function is another crucial factor for the overall performance. An objective function is used to compute the parameters in the mapping function. Popular signal-based distances include L1 \cite{pandey2019new} and L2 norms \cite{pandey2019new} and SI-SDR \cite{pandey2019new}. Recently, metric-based objective functions have been widely investigated; e.g., the objective functions that aim to enhance speech quality \cite{fu2021metricgan+}, speech intelligibility \cite{kolbaek2018monaural}, and automatic speech recognition \cite{lu20d_interspeech} performance have been derived. Although supervised SE methods have shown promising performance, noisy--clean paired training data may not always be available in real-world scenarios.        
	
	SE methods trained without the need for paired noisy--clean training data can be divided into two groups. The first group does not require any clean speech data, and this group of approaches is referred to as ``unsupervised SE methods." These methods are often based on assumptions regarding the characteristics of speech and noise signals. Traditional SE methods such as spectral subtraction \cite{boll1979suppression}, Wiener filtering \cite{lim1979enhancement}, and the minimum mean-square error of the spectral amplitude \cite{ephraim1985speech} belong to this unsupervised SE group. Recently, deep learning models have been used for unsupervised SE methods \cite{bie2021unsupervised, wang2020self,sivaraman2020self, fu2022metricgan, zezario2020self}. Generally, these approaches perform well in stationary noisy environments; however, when encountering non-stationary noises, they may produce subnormal results. The second group of approaches uses both noisy and clean speech data to prepare a better mapping function from the two domains, whereas the noisy and clean speech data are not paired. A well-known group of SE methods that uses unpaired noisy--clean speech data is the cycle-consistent SE method \cite{xiang2020parallel, meng2018cycle, yu2021two,yu2021cyclegan}. 
	Cycle-consistent learning was first proposed for unpaired image-to-image translations \cite{zhu2017unpaired} and has been successfully applied to voice conversion \cite{kaneko2019cyclegan}. When applied to unpaired SE methods, cycle-consistent learning adopts two generators and two discriminators, which can be viewed as a filter (converting noisy speech to clean speech) and a corresponding inverse filter (converting clean speech to noisy speech), and the two discriminators aim to distinguish real samples (real noisy speech and real clean speech) from fake ones forged by the generators. Only the filter (converting noisy speech to clean speech) is employed during testing to perform denoising. Let us take a closer look at the training stage of cycle-consistent SE methods; the filter is trained to convert noisy speech into clean speech, and the inverse filter is trained to convert clean speech into noisy speech. It is suitable for training the filter because multiple noisy speeches can be used as input. However, it may be suboptimal for estimating the inverse filter because no target noise is specified to generate, which consequently limits the achievable enhancement performance. 
	
	Herein, we eliminated this limitation using a series of experiments with a novel noise-informed training (NIT) approach along with a cycle-consistent generative adversarial network (CycleGAN), termed NIT-CycleGAN. The main feature of NIT-CycleGAN is that the target domain was specified when training the inverse filter. Specifically, a label of the target domain was appended to the acoustic features of the filter (and inverse filter) to specify the target domain of the output. The target domain label indicated whether the target was clean or of a particular noise type and was represented by a one-hot vector.
	Compared with the original CycleGAN, the proposed NIT-CycleGAN used additional information on the target domain during training. 
	During testing, we aimed to use a filter to convert noisy speech into clean speech. Thus, we simply set the target domain label as ``clean." The experimental results confirmed the effectiveness of the NIT approach. Moreover, it verified the decent generalization capability of NIT-CycleGAN in unseen noisy environments.
	
	
	\vspace{-0.2cm}\section{Related work}\vspace{-0.2cm}\label{sec:relwork}
	
	Given a set of noisy acoustic features, $\mathbf{y}$, the CycleGAN SE system adopted a filter to convert $\mathbf{y}$ into clean-like features, which were converted into the original $\mathbf{y}$ using an inverse filter. The filter and inverse filter of CycleGAN are trained using three losses, namely, adversarial, cycle-consistency, and identity-mapping losses. For a CycleGAN, generators $G^{\mathbf{Y\rightarrow S}}$ and $G^{\mathbf{S\rightarrow Y}}$ performed filter and inverse filter processing, respectively. Specifically, the generator $G^{\mathbf{Y\rightarrow S}}$ was applied to convert $\mathbf{y}$ into clean-like speech ${\mathbf{s}}$; conversely, the generator $G^{\mathbf{S\rightarrow Y}}$ performed a reverse process that converted $\mathbf{s}$ into ${\mathbf{y}}$. 
	Two discriminators, $D^{\mathbf{Y}}$ and $D^{\mathbf{S}}$, were used to identify whether the generated features were distinguishable from the real ones.
	
	\vspace{-0.2cm}\subsection{Cycle-consistency loss}\vspace{-0.2cm}
	The cycle-consistency loss function $\mathcal{L}_\mathrm{cyc}$ is expressed in Eq. \eqref{eq:cyc}.
	\begin{equation}\label{eq:cyc}
		\begin{scriptsize}
			\begin{aligned}
				\mathcal{L}_\mathrm{cyc}&\{G^{\mathbf{S\rightarrow Y}}, G^{\mathbf{Y\rightarrow S}}, \mathbf{s}, \mathbf{y}\}=\mathbb{E}_{\mathbf{s}\sim  P_{\mathbf{S}}}\{\| G^{\mathbf{Y\rightarrow S}}(G^{\mathbf{S\rightarrow Y}}(\mathbf{s}))-\mathbf{s}\|_1\}\\
				&+\mathbb{E}_{\mathbf{y}\sim  P_{\mathbf{Y}}}\{\|G^{\mathbf{S\rightarrow Y}}(G^{\mathbf{Y\rightarrow S}}(\mathbf{y}))-\mathbf{y}\|_1\},
			\end{aligned}
		\end{scriptsize}
	\end{equation}
	where $\mathbb{E}_{*}\{\cdot\}$ denotes the expectation operation, and $P_{\mathbf{S}}$ and $P_{\mathbf{Y}}$ represents the clean and noisy domain distributions, respectively. Both generators were trained to minimize $\mathcal{L}_\mathrm{cyc}$ in Eq. \eqref{eq:cyc}. For CycleGAN SE, one goal of the cycle-consistency loss function was to regularize $G^{\mathbf{S\rightarrow Y}}$ and $G^{\mathbf{Y\rightarrow S}}$ to preserve acoustic content.
	
	\vspace{-0.2cm}\subsection{Adversarial loss}\vspace{-0.2cm}\label{sec:adv}
	To ensure that the generated speech $G^{\mathbf{Y\rightarrow S}}(\mathbf{y})$ and $G^{\mathbf{S\rightarrow Y}}(\mathbf{s})$ were indistinguishable from clean and noisy data, respectively, two adversarial losses were derived. 
	
	\vspace{-0.4cm}\subsubsection{First adversarial loss function}\vspace{-0.2cm}
	Regarding $G^{\mathbf{Y\rightarrow S}}$, the first adversarial loss function is expressed as Eq. \eqref{eq:advloss}.
	\begin{equation}\label{eq:advloss}
		\begin{scriptsize}
			\begin{aligned}
				\mathcal{L}_\mathrm{adv}^{1}&\{D^{\mathbf{S}}, G^{\mathbf{Y\rightarrow S}}, \mathbf{s}, \mathbf{y}\}=\mathbb{E}_{\mathbf{s}\sim  P_{\mathbf{S}}}\{log[ D^{\mathbf{S}}(\mathbf{s})]\}\\
				&+\mathbb{E}_{\mathbf{y}\sim  P_{\mathbf{Y}}}\{log[1-D^{\mathbf{S}}( G^{\mathbf{Y\rightarrow S}}(\mathbf{y}) )]\}.
			\end{aligned}
		\end{scriptsize}
	\end{equation}
	The minimax criterion was applied to $\mathcal{L}_\mathrm{adv}^{1}$ to iteratively optimize $G^{\mathbf{Y\rightarrow S}}$ and $D^{\mathbf{S}}$. 
	
	Similarly, the discriminator $D^{\mathbf{Y}}$ and generator $G^{\mathbf{S\rightarrow Y}}$ were updated by $\mathcal{L}_\mathrm{adv}^{1}\{D^{\mathbf{Y}}, G^{\mathbf{S\rightarrow Y}}, \mathbf{s}, \mathbf{y}\}$.
	
	\vspace{-0.2cm}\subsubsection{Second adversarial loss function}\vspace{-0.2cm}
	Eq. \eqref{eq:2ndadvloss} shows the second adversarial loss function.
	\begin{equation}\label{eq:2ndadvloss}
		\begin{aligned}
			\mathcal{L}_\mathrm{adv}^{2}&\{D^{\mathbf{S}}, G^{\mathbf{Y\rightarrow S}}, G^{\mathbf{S\rightarrow Y}}, \mathbf{s}\}=\mathbb{E}_{\mathbf{s}\sim  P_{\mathbf{S}}}\{log[ D^{\mathbf{S}}(\mathbf{s})]\}\\
			&+\mathbb{E}_{\mathbf{s}\sim  P_{\mathbf{S}}}\{log[1-D^{\mathbf{S}}( G^{\mathbf{Y\rightarrow S}}(G^{\mathbf{S\rightarrow Y}}(\mathbf{s})))]\}.
		\end{aligned}
	\end{equation}
	Based on $\mathcal{L}_\mathrm{adv}^{2} \{ D^{\mathbf{S}}, G^{\mathbf{Y\rightarrow S}}, G^{\mathbf{S\rightarrow Y}}, \mathbf{s} \}$, we updated
	$D^{\mathbf{S}}$.
	Similarly, $\mathcal{L}_\mathrm{adv}^{2}\{D^{\mathbf{Y}}, G^{\mathbf{Y\rightarrow S}}, G^{\mathbf{S\rightarrow Y}}, \mathbf{y}\}$ was used to update $D^{\mathbf{Y}}$. Notably, this loss was used in \cite{kaneko2019cyclegan} to improve VC performance.
	
	\vspace{-0.2cm}\subsection{Identity-mapping loss}\vspace{-0.2cm}
	Further, to improve the CycleGAN SE system, the identity--mapping loss function in Eq. \eqref{eq:idm} was adopted to ensure that the output of a generator was nearly identical to the input.	
	\begin{equation}\label{eq:idm}
		\begin{scriptsize}
			\begin{aligned}
				\mathcal{L}_\mathrm{idm}&\{G^{\mathbf{Y\rightarrow S}}, G^{\mathbf{S\rightarrow Y}}, \mathbf{s}, \mathbf{y}\}=\mathbb{E}_{\mathbf{s}\sim  P_{\mathbf{S}}}\{\| G^{\mathbf{Y\rightarrow S}}(\mathbf{s})-\mathbf{s}\|_1\}\\
				&+\mathbb{E}_{\mathbf{y}\sim  P_{\mathbf{Y}}}\{\|G^{\mathbf{S\rightarrow Y}}(\mathbf{y})-\mathbf{y}\|_1\}.
			\end{aligned}
		\end{scriptsize}	
	\end{equation}
	
	\vspace{-0.2cm}\section{Proposed Method}\vspace{-0.2cm}\label{sec:nacgan}
	
	The proposed NIT-CycleGAN SE system shared the same model architecture as the original CycleGAN system. Specifically, the NIT-CycleGAN SE system consisted of a noisy--clean generator, $G^{\mathbf{Y\rightarrow S}}$, a clean--noisy generator, $G^{\mathbf{S\rightarrow Y}}$, and two discriminators, $D^{\mathbf{S}}$ and $D^{\mathbf{Y}}$. The main idea of NIT-CycleGAN is to incorporate the target domain information during training. To this end, we adopted an auxiliary one-hot vector that indicated the target domain of the generated speech. For the noisy--clean generator, the one-hot vector indicated the output target to be ``clean.'' For the clean--noisy generator, the one-hot vector indicated the output target to be ``a specific noise type." Auxiliary vectors were used to provide target information to govern the generators to convert the source input into the specified target domain and serve as additional features for discriminators to identify differences between the original and generated samples. We concatenated the one-hot vector with each acoustic feature to form an extended feature in a frame-wise manner. In the following, we introduce the training and testing stages of NIT-CycleGAN.
	
	\vspace{-0.2cm}\subsection{Training stage}\vspace{-0.2cm}\label{sec:tr}
	Like CycleGAN, the NIT-CycleGAN SE system was trained with (NIT)-cycle-consistency, adversarial, and identity-mapping losses. 
	
	\vspace{-0.3cm}\subsubsection{Noise-informed-training cycle-consistency loss}\vspace{-0.2cm}
	Assume that there were $N$ different noise types in the training data and that the noise type for each training utterance was known, we formulated the auxiliary one-hot vector, $tn$, as an $(N+1)$-dimensional vector (one clean with $N$ noise types). In this one-hot vector, a single non-zero element corresponded to the target domain (a particular noise type). The one-hot vector $tn$ was appended with the acoustic feature $\mathbf{s}$, i.e., $\mathbf{s}_{tn}=[tn;\mathbf{s}]$. Then, this clean input vector $\mathbf{s}_{tn}$ was processed using generator $G^{\mathbf{S\rightarrow Y}}$ to create $\mathbf{y}'_{tn'}=[tn';\mathbf{y}']$. Notably, the $tn$ vector in the input was used to guide the generator to convert signal $\mathbf{s}$ into the designated noise domain. Next, we replaced $tn'$ from $\mathbf{y}'_{tn'}$ with a one-hot vector $tc$, where the non-zero element indicated that the target domain was a clean condition. Then, this result denoted as $\mathbf{y}'_{tc}$ was passed through $G^{\mathbf{Y\rightarrow S}}$  to generate an enhanced output $\mathbf{s}'_{tc'}$ with the appended vector $tc'$, where $\mathbf{s}'_{tc'}$ was expected to approximate the ground truth $\mathbf{s}_{tc}$. A similar procedure was conducted by passing a noisy vector $\mathbf{y}_{tc}$ to $G^{\mathbf{Y\rightarrow S}}$ and subsequently to $G^{\mathbf{S\rightarrow Y}}$ to obtain $\mathbf{y}'_{tn'}$. Thus, the NIT-cycle-consistency loss function can be written as:
	\begin{equation}\label{eq:nacyc}
		\begin{scriptsize}
			\begin{aligned}
				\mathcal{L}_\mathrm{nit-cyc}&\{G^{\mathbf{Y\rightarrow S}}, G^{\mathbf{S\rightarrow Y}}, \mathbf{s}_{tc}, \mathbf{s}_{tn}, \mathbf{y}_{tc},
				\mathbf{y}_{tn}\}=\\&\mathbb{E}_{\mathbf{s}\sim  P_{\mathbf{S}}}\{\| G^{\mathbf{Y\rightarrow S}}(G^{\mathbf{S\rightarrow Y}}(\mathbf{s}_{tn}))-\mathbf{s}_{tc}\|_1\}+\\&\mathbb{E}_{\mathbf{y}\sim  P_{\mathbf{Y}}}\{\|G^{\mathbf{S\rightarrow Y}}(G^{\mathbf{Y\rightarrow S}}(\mathbf{y}_{tc}))-\mathbf{y}_{tn}\|_1\}.
			\end{aligned}
		\end{scriptsize}
	\end{equation}
	
	\vspace{-0.4cm}\subsubsection{Noise-informed-training adversarial loss}\vspace{-0.2cm}
	The objective of the two generators was to generate acoustic features to deceive the corresponding discriminators. By contrast, the objective of a discriminator was to learn the detailed differences between real and generated samples to avoid being fooled by their generators. NIT-CycleGAN incorporated additional information about the target domain into the training process. Similar to the original CycleGAN SE system, two NIT-adversarial losses were used in NIT-CycleGAN.
	
	The NIT-adversarial loss with respect to $D^{\mathbf{S}}$ and $G^{\mathbf{Y\rightarrow S}}$ is expressed by Eq. \eqref{eq:naadv} in terms of Eq. \eqref{eq:advloss}.
	\begin{equation}\label{eq:naadv}
		\begin{scriptsize}
			\begin{aligned}
				\mathcal{L}_\mathrm{nit-adv}^{1}&\{D^{\mathbf{S}}, G^{\mathbf{Y\rightarrow S}}, \mathbf{s}_{tc}, \mathbf{y}_{tc}\} = \mathbb{E}_{\mathbf{s} \sim P_{\mathbf{S}}} \{ log [D^{\mathbf{S}} (\mathbf{s}_{tc})]\} \\ & +  \mathbb{E}_{\mathbf{y} \sim P_{\mathbf{Y}}} \{ log[1 - D^{\mathbf{S}}(G^{\mathbf{Y\rightarrow S}}(\mathbf{y}_{tc}))] \}.
			\end{aligned}
		\end{scriptsize}
	\end{equation}
	The two models, $D^{\mathbf{S}}$ and $G^{\mathbf{S\rightarrow Y}}$, were optimized based on $\mathcal{L}_\mathrm{nit-adv}^{1}$, in which $D^{\mathbf{S}}$ attempted to maximize $\mathcal{L}_\mathrm{nit-adv}^{1}$, whereas $G^{\mathbf{Y\rightarrow S}}$ did the opposite. Notably, $D^{\mathbf{S}}$ identified the difference between the clean vector ${\mathbf{s}}_{tc}$ and enhanced vector $\mathbf{s}'_{tc'}$ obtained by passing $\mathbf{y}_{tc}$ through the generator $G^{\mathbf{Y\rightarrow S}}$. The predicted vector $tc'$ in $\mathbf{s}'_{tc'}$ served as an auxiliary feature for $D^{\mathbf{S}}$. Similarly, the discriminator $D^{\mathbf{Y}}$ and generator $G^{\mathbf{S\rightarrow Y}}$ were optimized using the loss function $\mathcal{L}_\mathrm{nit-adv}^{1}\{D^{\mathbf{Y}}, G^{\mathbf{S\rightarrow Y}}, \mathbf{s}_{tn}, \mathbf{y}_{tn}\}$. 
	
	Next, Eq. \eqref{eq:2ndnaadv} describes the second NIT-adversarial loss with respect to $\mathbf{s}_{tc}$ and $\mathbf{s}_{tn}$.
	\begin{equation}\label{eq:2ndnaadv}
		\begin{scriptsize}
			\begin{aligned}
				\mathcal{L}_\mathrm{nit-adv}^{2}&\{D^{\mathbf{S}}, G^{\mathbf{Y\rightarrow S}}, G^{\mathbf{S\rightarrow Y}}, \mathbf{s}_{tn}, \mathbf{s}_{tc}\}= \mathbb{E}_{\mathbf{s}\sim  P_{\mathbf{S}}}\{log[ D^{\mathbf{S}}(\mathbf{s}_{tc})]\}\\
				&+\mathbb{E}_{\mathbf{s}\sim  P_{\mathbf{S}}}\{log[1-D^{\mathbf{S}}( G^{\mathbf{Y\rightarrow S}}(G^{\mathbf{S\rightarrow Y}}(\mathbf{s}_{tn})))]\}.
			\end{aligned}
		\end{scriptsize}
	\end{equation}
	Here, the second NIT-adversarial loss in Eq. \eqref{eq:2ndnaadv} was used to optimize the discriminator $D^{\mathbf{S}}$. Similarly, $D^{\mathbf{Y}}$ was estimated using 
	$\mathcal{L}_\mathrm{nit-adv}^{2}\{D^{\mathbf{Y}}, G^{\mathbf{S\rightarrow Y}}, G^{\mathbf{Y\rightarrow S}}, \mathbf{y}_{tc}, \mathbf{y}_{tn}\}$ applied to the minimax criterion.
	
	\vspace{-0.3cm}\subsubsection{Noise-informed-training identity-mapping loss}\vspace{-0.2cm}
	The NIT-identity-mapping loss function is expressed as: 
	\begin{equation}\label{eq:naidm}
		\begin{scriptsize}
			\begin{aligned}
				\mathcal{L}_\mathrm{nit-idm}&\{G^{\mathbf{Y\rightarrow S}}, G^{\mathbf{S\rightarrow Y}}, \mathbf{s}_{tc}, \mathbf{y}_{tn}\} = \mathbb{E}_{\mathbf{s} \sim P_{\mathbf{S}}} \{ \| G^{\mathbf{Y\rightarrow S}}(\mathbf{s}_{tc}) - \mathbf{s}_{tc} \|_1 \} \\ & + \mathbb{E}_{\mathbf{y} \sim P_{\mathbf{Y}}} \{ \| G^{\mathbf{S\rightarrow Y}}(\mathbf{y}_{tn}) -\mathbf{y}_{tn} \|_1 \}.
			\end{aligned}
		\end{scriptsize}
	\end{equation}
	From the NIT-identity-mapping loss function, the predicted auxiliary vector at the generator output was expected to be identical to that of the model input.
	
	\begin{figure}[!t]
		\begin{center}
			\includegraphics[width=0.75\columnwidth]{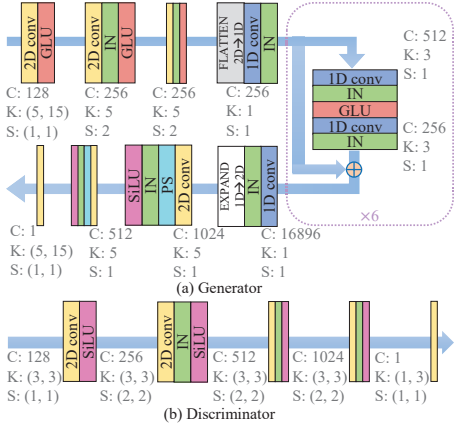}\vspace{-0.3cm}
			\caption{Model architectures of (a) generator and (b) discriminator in both CycleGAN and NIT-CycleGAN SE systems. The notations ``C,'' ``K,'' and ``S'' represent the output channels, kernel size, and stride of a convolution layer, respectively.\vspace{-0.7cm}}\label{fig:mdlconfig}
		\end{center}
	\end{figure}
	
	\vspace{-0.2cm}\subsection{Testing stage}\vspace{-0.2cm}
	In the testing stage, since the objective was to generate clean-like enhanced speech, we assigned the one-hot vector to be ``clean'' as the target domain, namely specifying $tc$ as an auxiliary input. This one-hot vector was appended to each frame of the noisy acoustic features to form extended features. Then, the extended features were passed to the generator, producing enhanced features ($\mathbf{s}'$ and $tc'$). The enhanced features were combined with the phase of the noisy speech input and converted into time-domain enhanced speech signals.
	
	\vspace{-0.2cm}\section{Experiment and Analysis}\label{sec:exp}
	\vspace{-0.2cm}\subsection{Experimental setup}\vspace{-0.2cm}
	We assessed the proposed NIT-CycleGAN SE system on the Taiwan Mandarin Hearing in Noise Test dataset \cite{huang2005development}. The dataset contains 2,560 clean speech utterances spoken by four male and four female speakers and was recorded at a 16-kHz sampling rate. Among these utterances, we selected those spoken by three male and three female speakers to prepare the training sets. Two training sets were prepared. For the first training set, termed ``Cat-L,'' 1,194 clean utterances were contaminated by five noises (i.e., dwashing, npark, straffic, pcafeter, and tbus); accordingly, $N=5$, as Sec. \ref{sec:tr} describes, at signal-to-noise ratios (SNRs) of $-5$, $0$, and $5$ dB. Thus, we prepared 17,910 noisy-clean training pairs. For the second training set, we split 1,194 clean utterances into two parts, and the contents of the utterances in these two parts are different. Based on the first 597 clean utterances, we used the same five noise types to generate 8,955 ($597\times5\times3$) noisy utterances at $-5$, $0$, and $5$ dB SNRs. The remaining 597 clean recordings along with the 8,955 noisy utterances were combined to form the second training set, termed ``Cat-S.'' Notably, the data size of ``Cat-L'' is larger than that of ``Cat-S,'' and the contents of clean and noisy utterances are different for ``Cat-S.'' We assessed the performance of CycleGAN and NIT-CycleGAN SE systems using both ``Cat-L'' and ``Cat-S'' training sets. The testing set was prepared using 240 clean utterances spoken by the other male and female speakers. Each utterance in the testing corpus was deteriorated by the five noise types used in the training data (matched conditions) and seven unseen noise types (tmetro, tcar, spsquare, pstation, presto, ooffice, and nfield) at SNRs of -5, 0, and 5 dB (mismatched conditions). Consequently, 8,640 noisy utterances were used for the evaluation. Here, all the noise sources used were collected from DEMAND \cite{thiemann2013demand}.
	
	The frame size and hop length were set at 32 ms and 16 ms, respectively, to apply the short-time Fourier transform to convert speech waveforms into 257-dimensional spectral features. Fig. \ref{fig:mdlconfig} shows the detailed model architectures for the generators and discriminators, where GLU, SiLU, IN, and PS represent the gated linear units, sigmoid linear units, instance normalization, and pixel shuffling processes, respectively. All networks were trained for 600 epochs using the adaptive moment estimation (Adam) optimizer with beta values of 0.5 and 0.999. The learning rates were 0.0002 and 0.0001 for the generators and discriminators, respectively.
	
	Four objective metrics were used to assess the proposed system: (1) perceptual evaluation of speech quality (PESQ) \cite{rix2001perceptual}, (2) mean opinion score (MOS) prediction of speech signal distortion (CSIG) \cite{hu2007evaluation}, (3) MOS prediction of the intrusiveness of background noise (CBAK) \cite{hu2007evaluation}, and (4) MOS prediction of the overall effect (COVL) \cite{hu2007evaluation}. The score range of PESQ is $[-0.5, 4.5]$, whereas those of CSIG, CBAK, and COVL are $[1, 5]$. Higher scores for PESQ, CSIG, CBAK, and COVL indicate better sound quality, lower signal distortion, residual noise, and overall rating, respectively.\vspace{-0.1cm}
	
	\vspace{-0.2cm}\subsection{The Cat-L training set}\vspace{-0.2cm}\label{sec:sup}
	In this subsection, both CycleGAN and NIT-CycleGAN models were trained on the ``Cat-L'' training set and assessed on the same testing set. Table \ref{tab:match_Cat_I} presents the average CBAK, COVL, CSIG, and PESQ scores for noisy speech (denoted as ``Noisy''), enhanced utterances using CycleGAN (denoted as ``CycleGAN--L''), and NIT-CycleGAN (denoted as ``NIT-CycleGAN--L''). From the table, CycleGAN--L and NIT-CycleGAN--L demonstrate improved scores over Noisy, whereas NIT-CycleGAN--L outperformed CycleGAN--L in terms of CBAK, COVL, and CSIG evaluation metrics. The results suggest that by applying the NIT approach, CycleGAN could improve its performance in matched testing conditions.
	
	Next, we list the average CBAK, COVL, CSIG, and PESQ scores of Noisy, CycleGAN--L, and NIT-CycleGAN--L assessed under mismatched testing conditions in Table \ref{tab:mismatch_Cat_I}. From the table, we observe that all evaluation scores of the CycleGAN--L and NIT-CycleGAN--L approaches outperformed those from Noisy, confirming the effectiveness of the CycleGAN-based architecture of SE on the unseen noisy types. Also, NIT-CycleGAN--L yielded higher scores than CycleGAN--L in terms of the CBAK, COVL, and CSIG evaluation metrics, again confirming the effectiveness of the NIT approach for CycleGAN with unpaired noisy--clean training data. 
	
	\begin{table}[!b]
		\vspace{-0.3cm}\caption{Evaluation scores for Noisy, CycleGAN--L, and NIT-CycleGAN--L under matched conditions.}\vspace{-0.3cm}\label{tab:match_Cat_I}
		\centering
		\begin{tabularx}{\columnwidth}{>{\centering}m{2.4cm}|>{\centering}m{0.97cm}>{\centering}m{0.97cm}>{\centering}m{0.97cm}>{\centering\arraybackslash}X}
			\toprule
			\hline
			& CSIG & CBAK & COVL &PESQ \\
			\hline
			Noisy     & 2.986& 1.916 & 2.268 & 2.353 \\
			CycleGAN--L  & 3.312& 2.456 & 2.633 & \textbf{2.721} \\
			NIT-CycleGAN--L & \bf{3.371} & \bf{2.469} & \bf{2.667} & 2.718\\
			\hline
			\bottomrule
		\end{tabularx}
	\end{table}		
	\begin{table}[!b]
		\caption{Evaluation scores for Noisy, CycleGAN--L, and NIT-CycleGAN--L on mismatched conditions.}\vspace{-0.3cm}\label{tab:mismatch_Cat_I}
		\centering
		\begin{tabularx}{\columnwidth}{>{\centering}m{2.4cm}|>{\centering}m{0.97cm}>{\centering}m{0.97cm}>{\centering}m{0.97cm}>{\centering\arraybackslash}X}
			\toprule
			\hline
			& CSIG & CBAK & COVL &PESQ \\
			\hline
			Noisy    & 2.854& 1.810 & 2.071 & 2.249\\
			CycleGAN--L & 3.249& 2.374 & 2.553 & \textbf{2.650}\\
			NIT-CycleGAN--L & \bf{3.308}& \bf{2.389} & \bf{2.588}& 2.648 \\
			\hline
			\bottomrule
		\end{tabularx}
	\end{table}	
	\begin{table}[!b]
		\caption{Evaluation scores for Noisy, CycleGAN--S, and NIT-CycleGAN--S under all testing conditions.}\vspace{-0.3cm}\label{tab:match_Cat_II}
		\centering
		\begin{tabularx}{\columnwidth}{>{\centering}m{2.4cm}|>{\centering}m{0.97cm}>{\centering}m{0.97cm}>{\centering}m{0.97cm}>{\centering\arraybackslash}X}
			\toprule
			\hline
			& CSIG & CBAK & COVL &PESQ \\
			\hline
			Noisy    & 2.909&1.854&2.153 & 2.293 \\
			CycleGAN--S  &3.089&2.260&2.418 &2.614 \\
			NIT-CycleGAN--S &\textbf{3.335}&\textbf{2.400}&\textbf{2.602}&\textbf{2.679} \\
			\hline
			\bottomrule
		\end{tabularx}
	\end{table}
	
	\vspace{-0.2cm}\subsection{The Cat-S training set}\vspace{-0.2cm}
	Next, we investigate the performance of CycleGAN and NIT-CycleGAN SE systems, trained on ``Cat-S'' and assessed on matched and mismatched testing sets. Table \ref{tab:match_Cat_II} lists the average CBAK, COVL, and CSIG scores for Noisy, enhanced speech by CycleGAN (``CycleGAN--S'') and by NIT-CycleGAN (``NIT-CycleGAN--S'') for the 12 noise conditions. The table shows that NIT-CycleGAN--S outperformed CycleGAN--S in all evaluation metrics, confirming the advantage of the NIT approach for CycleGAN on the SE task. 
	
	

	\vspace{-0.2cm}\subsection{Discussion}\vspace{-0.2cm}
	\subsubsection{Noise level}
\begin{figure}[!t]
		\begin{center}
			\includegraphics[width=\columnwidth]{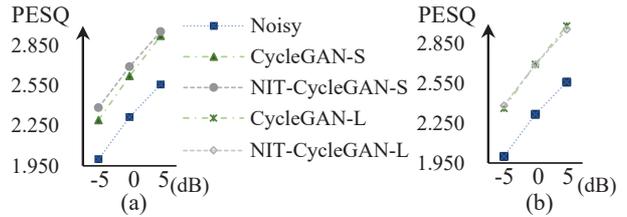}
			\vspace{-0.3cm}\caption{PESQ values of (a) Noisy, CycleGAN--S and NIT-CycleGAN--S along with (b) Noisy, CycleGAN--L and NIT-CycleGAN--L at $-5$, $0$, and $5$ dB SNRs. \vspace{-0.4cm}}\label{fig:snr}
		\end{center}
	\end{figure}

\vspace{-0.2cm}
	In Fig. \ref{fig:snr}, we compared (a) CycleGAN--S with NIT-CycleGAN--S and (b) CycleGAN--L with NIT-CycleGAN--L on different SNR levels. We report the performances of these SE systems in terms of the PESQ metric. Fig. \ref{fig:snr} (a) and (b) show the noisy results as the baseline. From both figures, all CycleGAN-based SE systems generated utterances with higher quality than Noisy under each SNR condition. Additionally, along the \textit{x}-axis (dB), the results of NIT-CycleGAN--S are consistently higher than those of CycleGAN--S in Fig. \ref{fig:snr} (a), and comparable PESQ values between NIT-CycleGAN--L and CycleGAN--L are observed in Fig. \ref{fig:snr} (b). When comparing the results of ``Cat-S'' and ``Cat-L'' training sets, we note the advantage brought by the NIT approach was clearer when a smaller training set was available.
	
	\begin{figure}[!t]
		\begin{center}
			\includegraphics[width=0.95\columnwidth]{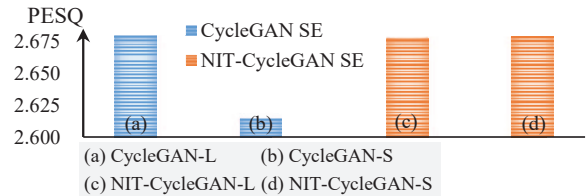}
			\vspace{-0.3cm}\caption{Mean PESQ values for (a) CycleGAN--L, (b) CycleGAN--S, (c) NIT-CycleGAN--L, and (d) NIT-CycleGAN--S on all noisy conditions. The blue and red bars represent the quality scores obtained individually from the CycleGAN- and NIT-CycleGAN-based SE systems, respectively.\vspace{-0.7cm}}\label{fig:sws}
		\end{center}
	\end{figure}
	
	
	\subsubsection{Generalization capability}\vspace{-0.2cm}
	We then report the generalization capability of the proposed NIT-CycleGAN. Fig. \ref{fig:sws} shows (a) CycleGAN--L, (b) CycleGAN--S, (c) NIT-CycleGAN--L, and (d) NIT-CycleGAN--S, wherein the \textit{x}-axis represents the applied SE systems and the \textit{y}-axis being the PESQ values. For each SE system, the quality scores were calculated by averaging the values for all the noise conditions. In Fig. \ref{fig:sws}, the quality difference between CycleGAN--L and CycleGAN--S is larger than that between NIT-CycleGAN--L and NIT-CycleGAN--S. The difference between NIT-CycleGAN--L and NIT-CycleGAN--S is marginal. The results suggest that the NIT approach could increase the generalization capability of CycleGAN, which enabled it to achieve good performance even with a small amount of training data.
	
	\vspace{-0.2cm}\section{Conclusion}\vspace{-0.2cm}\label{sec:summary}
	This study discussed a potential limitation of the CycleGAN-based SE system and proposed a novel NIT-CycleGAN. The experimental results showed that by incorporating the information of the target domain during training, NIT-CycleGAN could yield improved performance over CycleGAN for larger and smaller training sets under both matched and mismatched testing conditions. Additionally, we promoted CycleGAN SE by concatenating target-domain indicators with noisy input without changing the model architectures. The results verified the advantage of the increased generalization capability using the NIT approach. In the future, we will explore the incorporation of various target attributes, such as SNR and speakers, to further improve the CycleGAN SE. Meanwhile, we will explore the use of other model architectures for NIT-CycleGAN.
	
	\bibliographystyle{IEEEtran}
	\bibliography{refs}

\begin{thebibliography}{10}
\providecommand{\url}[1]{#1}
\csname url@samestyle\endcsname
\providecommand{\newblock}{\relax}
\providecommand{\bibinfo}[2]{#2}
\providecommand{\BIBentrySTDinterwordspacing}{\spaceskip=0pt\relax}
\providecommand{\BIBentryALTinterwordstretchfactor}{4}
\providecommand{\BIBentryALTinterwordspacing}{\spaceskip=\fontdimen2\font plus
\BIBentryALTinterwordstretchfactor\fontdimen3\font minus
  \fontdimen4\font\relax}
\providecommand{\BIBforeignlanguage}[2]{{%
\expandafter\ifx\csname l@#1\endcsname\relax
\typeout{** WARNING: IEEEtran.bst: No hyphenation pattern has been}%
\typeout{** loaded for the language `#1'. Using the pattern for}%
\typeout{** the default language instead.}%
\else
\language=\csname l@#1\endcsname
\fi
#2}}
\providecommand{\BIBdecl}{\relax}
\BIBdecl

\bibitem{chao2021cross}
F.-A. Chao, J.-w. Hung, and B.~Chen, ``Cross-domain single-channel speech
  enhancement model with bi-projection fusion module for noise-robust {ASR},''
  in \emph{Proc. ICME}, 2021, pp. 1--6.

\bibitem{hartmann2013direct}
W.~Hartmann, A.~Narayanan, E.~Fosler-Lussier, and D.~Wang, ``A direct masking
  approach to robust {ASR},'' \emph{IEEE/ACM Transactions on Audio, Speech, and
  Language Processing}, vol.~21, no.~10, pp. 1993--2005, 2013.

\bibitem{mohammadiha2013supervised}
N.~Mohammadiha, P.~Smaragdis, and A.~Leijon, ``Supervised and unsupervised
  speech enhancement using nonnegative matrix factorization,'' \emph{IEEE/ACM
  Transactions on Audio, Speech, and Language Processing}, vol.~21, no.~10, pp.
  2140--2151, 2013.

\bibitem{kim2018unpaired}
G.~Kim, H.~Lee, B.-K. Kim, S.-H. Oh, and S.-Y. Lee, ``Unpaired speech
  enhancement by acoustic and adversarial supervision for speech recognition,''
  \emph{IEEE Signal Processing Letters}, vol.~26, no.~1, pp. 159--163, 2018.

\bibitem{neri2021unsupervised}
J.~Neri, R.~Badeau, and P.~Depalle, ``Unsupervised blind source separation with
  variational auto-encoders,'' in \emph{Proc. EUSIPCO}, 2021, pp. 311--315.

\bibitem{venkataramani2019style}
S.~Venkataramani, E.~Tzinis, and P.~Smaragdis, ``A style transfer approach to
  source separation,'' in \emph{Proc. WASPAA}, 2019, pp. 170--174.

\bibitem{zhao2018two}
Y.~Zhao, Z.-Q. Wang, and D.~Wang, ``Two-stage deep learning for
  noisy-reverberant speech enhancement,'' \emph{IEEE/ACM transactions on audio,
  speech, and language processing}, vol.~27, no.~1, pp. 53--62, 2018.

\bibitem{xu2014dynamic}
Y.~Xu, J.~Du, L.-R. Dai, and C.-H. Lee, ``Dynamic noise aware training for
  speech enhancement based on deep neural networks,'' in \emph{Proc.
  INTERSPEECH}, 2014, pp. 2670--2674.

\bibitem{lu2013speech}
X.~Lu, Y.~Tsao, S.~Matsuda, and C.~Hori, ``Speech enhancement based on deep
  denoising autoencoder,'' in \emph{Proc. INTERSPEECH}, 2013, pp. 436--440.

\bibitem{pandey2019tcnn}
A.~Pandey and D.~Wang, ``{TCNN}: Temporal convolutional neural network for
  real-time speech enhancement in the time domain,'' in \emph{Proc. ICASSP},
  2019, pp. 6875--6879.

\bibitem{sun2017multiple}
L.~Sun, J.~Du, L.-R. Dai, and C.-H. Lee, ``Multiple-target deep learning for
  lstm-rnn based speech enhancement,'' in \emph{Proc. HSCMA}, 2017, pp.
  136--140.

\bibitem{kim2020t}
J.~Kim, M.~El-Khamy, and J.~Lee, ``{T-GSA}: Transformer with gaussian-weighted
  self-attention for speech enhancement,'' in \emph{Proc. ICASSP}, 2020, pp.
  6649--6653.

\bibitem{pandey2019new}
A.~Pandey and D.~Wang, ``A new framework for cnn-based speech enhancement in
  the time domain,'' \emph{IEEE/ACM Transactions on Audio, Speech, and Language
  Processing}, vol.~27, no.~7, pp. 1179--1188, 2019.

\bibitem{fu2021metricgan+}
S.-W. Fu, C.~Yu, T.-A. Hsieh, P.~Plantinga, M.~Ravanelli, X.~lu, and Y.~Tsao,
  ``Metricgan+: An improved version of metricgan for speech enhancement,'' in
  \emph{Proc. INTERSPEECH}, 2021, pp. 201--205.

\bibitem{kolbaek2018monaural}
M.~Kolb{\ae}k, Z.-H. Tan, and J.~Jensen, ``Monaural speech enhancement using
  deep neural networks by maximizing a short-time objective intelligibility
  measure,'' in \emph{Proc. ICASSP}, 2018, pp. 5059--5063.

\bibitem{lu20d_interspeech}
Y.-J. Lu, C.-F. Liao, X.~Lu, J.~weih Hung, and Y.~Tsao, ``{Incorporating Broad
  Phonetic Information for Speech Enhancement},'' in \emph{Proc. INTERSPEECH},
  2020, pp. 2417--2421.

\bibitem{boll1979suppression}
S.~Boll, ``Suppression of acoustic noise in speech using spectral
  subtraction,'' \emph{IEEE Transactions on Acoustics, Speech and Signal
  Processing}, vol.~27, no.~2, pp. 113--120, 1979.

\bibitem{lim1979enhancement}
J.~S. Lim and A.~V. Oppenheim, ``Enhancement and bandwidth compression of noisy
  speech,'' \emph{Proceedings of the IEEE}, vol.~67, no.~12, pp. 1586--1604,
  1979.

\bibitem{ephraim1985speech}
Y.~Ephraim and D.~Malah, ``Speech enhancement using a minimum mean-square error
  log-spectral amplitude estimator,'' \emph{IEEE Transactions on Acoustics,
  Speech and Signal Processing}, vol.~33, no.~2, pp. 443--445, 1985.

\bibitem{bie2021unsupervised}
X.~Bie, S.~Leglaive, X.~Alameda-Pineda, and L.~Girin, ``Unsupervised speech
  enhancement using dynamical variational auto-encoders,'' \emph{arXiv preprint
  arXiv:2106.12271}, 2021.

\bibitem{wang2020self}
Y.-C. Wang, S.~Venkataramani, and P.~Smaragdis, ``Self-supervised learning for
  speech enhancement,'' \emph{arXiv preprint arXiv:2006.10388}, 2020.

\bibitem{sivaraman2020self}
A.~Sivaraman and M.~Kim, ``Self-supervised learning from contrastive mixtures
  for personalized speech enhancement,'' \emph{arXiv preprint
  arXiv:2011.03426}, 2020.

\bibitem{fu2022metricgan}
S.-W. Fu, C.~Yu, K.-H. Hung, M.~Ravanelli, and Y.~Tsao, ``{MetricGAN-U}:
  Unsupervised speech enhancement/dereverberation based only on
  noisy/reverberated speech,'' in \emph{Proc. ICASSP}, 2022, pp. 7412--7416.

\bibitem{zezario2020self}
R.~E. Zezario, T.~Hussain, X.~Lu, H.-M. Wang, and Y.~Tsao, ``Self-supervised
  denoising autoencoder with linear regression decoder for speech
  enhancement,'' in \emph{Proc. ICASSP}, 2020, pp. 6669--6673.

\bibitem{xiang2020parallel}
Y.~Xiang and C.~Bao, ``A parallel-data-free speech enhancement method using
  multi-objective learning cycle-consistent generative adversarial network,''
  \emph{IEEE/ACM Transactions on Audio, Speech, and Language Processing},
  vol.~28, pp. 1826--1838, 2020.

\bibitem{meng2018cycle}
Z.~Meng, J.~Li, Y.~Gong \emph{et~al.}, ``Cycle-consistent speech enhancement,''
  \emph{arXiv preprint arXiv:1809.02253}, 2018.

\bibitem{yu2021two}
G.~Yu, Y.~Wang, H.~Wang, Q.~Zhang, and C.~Zheng, ``A two-stage complex network
  using cycle-consistent generative adversarial networks for speech
  enhancement,'' \emph{Speech Communication}, vol. 134, pp. 42--54, 2021.

\bibitem{yu2021cyclegan}
G.~Yu, Y.~Wang, C.~Zheng, H.~Wang, and Q.~Zhang, ``Cyclegan-based non-parallel
  speech enhancement with an adaptive attention-in-attention mechanism,''
  \emph{arXiv preprint arXiv:2107.13143}, 2021.

\bibitem{zhu2017unpaired}
J.-Y. Zhu, T.~Park, P.~Isola, and A.~A. Efros, ``Unpaired image-to-image
  translation using cycle-consistent adversarial networks,'' in \emph{Proc.
  ICCV}, 2017, pp. 2223--2232.

\bibitem{kaneko2019cyclegan}
T.~Kaneko, H.~Kameoka, K.~Tanaka, and N.~Hojo, ``Cyclegan-vc2: Improved
  cyclegan-based non-parallel voice conversion,'' in \emph{Proc. ICASSP}, 2019,
  pp. 6820--6824.

\bibitem{huang2005development}
M.-W. Huang, ``Development of {Taiwan Mandarin} hearing in noise test,''
  \emph{Master thesis, Department of speech language pathology and audiology,
  National Taipei University of Nursing and Health science}, 2005.

\bibitem{thiemann2013demand}
J.~Thiemann, N.~Ito, and E.~Vincent, ``Demand: a collection of multi-channel
  recordings of acoustic noise in diverse environments,'' in \emph{Proc.
  Meetings Acoust}, 2013, pp. 1--6.

\bibitem{rix2001perceptual}
A.~W. Rix, J.~G. Beerends, M.~P. Hollier, and A.~P. Hekstra, ``Perceptual
  evaluation of speech quality ({PESQ})-a new method for speech quality
  assessment of telephone networks and codecs,'' in \emph{Proc. ICASSP}, 2001,
  pp. 749--752.

\bibitem{hu2007evaluation}
Y.~Hu and P.~C. Loizou, ``Evaluation of objective quality measures for speech
  enhancement,'' \emph{IEEE Transactions on audio, speech, and language
  processing}, vol.~16, no.~1, pp. 229--238, 2007.

\end{thebibliography}
	
\end{document}